# Explaining Planetary-Rotation Periods Using an Inductive Method

Gizachew Tiruneh, Ph. D., Department of Political Science, University of Central Arkansas
(First submitted on June 19th, 2009; last updated on June 15th, 2011)

This paper uses an inductive method to investigate the factors responsible for variations in planetary-rotation periods. I began by showing the presence of a correlation between the masses of planets and their rotation periods. Then I tested the impact of planetary radius, acceleration, velocity, and torque on rotation periods. I found that velocity, acceleration, and radius are the most important factors in explaining rotation periods. The effect of mass may be rather on influencing the size of the radii of planets. That is, the larger the mass of a planet, the larger its radius. Moreover, mass does also influence the strength of the rotational force, torque, which may have played a major role in setting the initial constant speeds of planetary rotation.

Many astronomers believe that planetary rotation is influenced by angular momentum and related phenomena, which may have occurred during the formation of the solar system (Alfven, 1976; Safronov, 1995; Artemev and Radzievskii, 1995; Seeds, 2001; Balbus, 2003). There is, however, not much identifiable recent research on planetary-rotation periods. Hughes' (2003) review of the literature reveals that much research is needed to understand the phenomenon of planetary spin. Some of the assumptions that scholars have made in the past, according to Hughes (2003), include that planetary spin is a function of the gaseous or terrestrial nature of planets, that there is a power law relationship between planetary spin angular momentum and planetary mass, that there is a relationship between planets' escape velocities and their spin rates, and that mass-independent spin periods can be obtained if one posits that a planetary formation process is governed by the balance between gravitational and centrifugal forces at the planetary equator. However, Hughes (2003) seems to imply that theoretical equations that scholars have proposed to measure rotation periods do not, for the most part, seem to yield identical results to observed planetary-spin rates.

One recent study, Park (2008), attempts to relate planetary-rotation period with torques exerted by a planet on its moon(s) and vice versa. Park (2008) claims that his calculations provide close estimate of planetary-rotation periods. For instance, he finds 23.65, 24.623, 9.925, 10.657, 17.245, and 16.11 hours for Earth, Mars, Jupiter, Saturn, Uranus, and Neptune, respectively. But his estimates also provide 45.5 days for the rotation period of the Sun. Although the estimate of Sun's rotation period seems very high (compared to its observed spin period of 25 days), Park (2008: 19) argues, "Considering $10^{20}$ order of astronomical numbers involved in the equations, the calculated number is close enough to observation…" Furthermore, Park (2008) did not calculate the rotation periods of Mercury and Venus, implying that his equations required the presence of moons orbiting planets.

This paper uses an inductive method to investigate the factors responsible for variations in planetary-rotation periods.[1] The study is inductive-based because it is predicated on an observation of empirical regularity in planetary data, not on a hypothesis deduced from a general theory. Specifically, I begin by showing a correlation between the masses of planets and their rotation periods. Then I examine the impact of planetary acceleration, velocity, radius, and torque on rotational periods. I find that velocity, acceleration, and radius are the most important factors in explaining rotation periods. The effect of mass may be rather on influencing the strength of the rotational force, torque; the latter is heavily dependent on the former. Mass does also influence the size of the radii of planets; that is, the larger the mass of a planet, the larger its radius.

---

[1] The sidereal rotation period of a planet is not always identical to its synodic period. The sidereal period, which is considered the true orbital period of a planet, is the time it takes a planet to make one-full revolution around the Sun. The synodic period is the time it takes a planet to reappear at the same point in the sky relative to the Sun. For instance, Earth's sidereal rotation period is 29.93, but its synodic period is 24 hours.

**Observing a Correlation between the Masses of Planets and their Rotation Periods**

Data on rotation, revolution, and other characteristics of planets have been collected using radar and similar technologies since the middle of the twentieth century. Upon inspection of such observational data, I noticed that the masses of the planets, with the exception of Mercury and Venus, were related to their rotation periods.

Table 1 shows that as the planetary masses increase, planetary-rotation periods decrease. The exceptions to this regularity are Mercury and Venus. Excluding Mercury and Venus, for being outliers, I plotted the foregoing relationship on a graph. As shown in Graph 1, I find that not only is the relationship between planetary masses and rotational periods negative, it is also logarithmic.

**Table 1: Planetary Mass and Rotation Periods**

| Planets | Mass (kg) | Sidereal Rotation Period (hours) |
|---|---|---|
| Pluto | $1.20 \times 10^{22}$ | 153.35 |
| Mercury | $3.31 \times 10^{23}$ | 1407.07 |
| Mars | $0.64 \times 10^{24}$ | 24.62 |
| Venus | $4.87 \times 10^{24}$ | 5839.20 |
| Earth | $5.98 \times 10^{24}$ | 23.93 |
| Uranus | $8.69 \times 10^{25}$ | 17.23 |
| Neptune | $1.03 \times 10^{26}$ | 16.05 |
| Saturn | $5.69 \times 10^{26}$ | 10.23 |
| Jupiter | $1.90 \times 10^{27}$ | 9.92 |

**Graph 1: Planetary Mass and Rotation Period\***

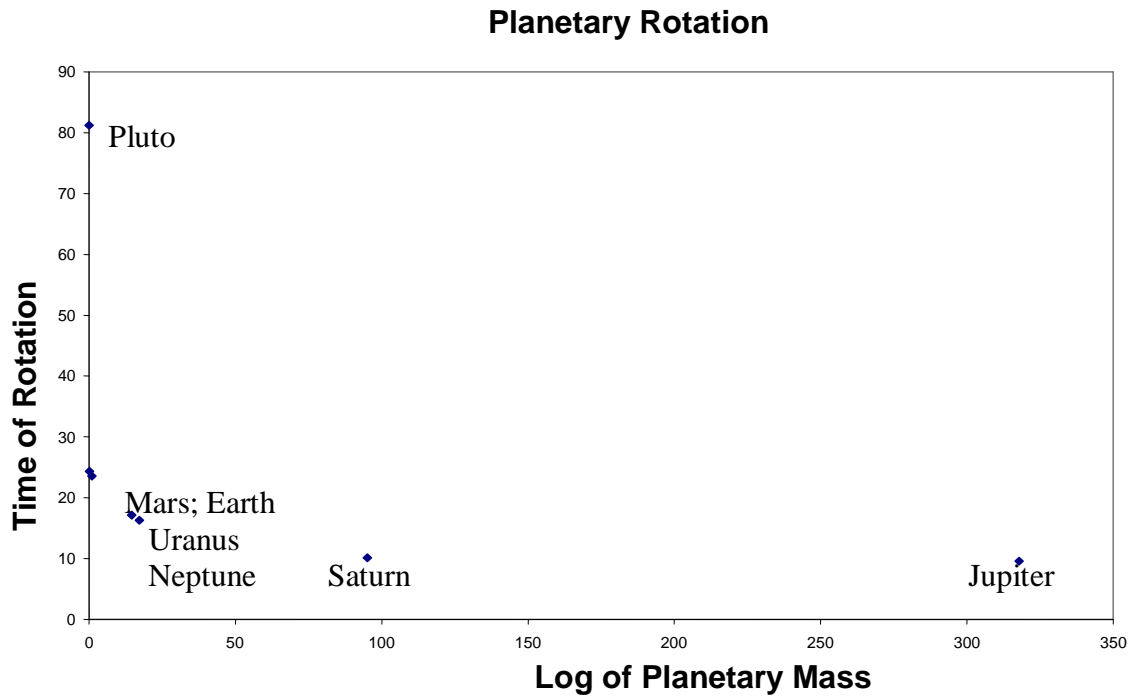

\* Note: Masses are relative to Earths, where Earth = 1.0; Rotation periods are in hours

Moreover, despite the small size of the cases, I ran a regression analysis to see the strength of the relationship between planetary masses and their rotational periods.[2] Given that Pluto is not considered a planet anymore, I excluded it from the following and the rest of the analyses. In addition, to avoid data skewness, I have omitted Mercury and Venus from subsequent regression analyses. They will be, however, included in data shown in tabular form. I used an inverse logarithmic model to describe the relationship. I logged the masses of the planets to make them consistent with the logarithmic model. The model is given by:

$$Y = a - b \ln X + e \qquad (1)$$

---

[2] Hughes (2003) shows that previous research has conducted regression analysis relating the logarithm of planetary mass to the logarithm of planetary spin angular momentum. Such analyses are not, however, identical to mine since my regression analysis relates the logarithm of planetary mass directly to (and to unlogged) planetary-rotation periods.

where Y = planetary-rotation periods, a = the Y-intercept, b = the slope, ln X = logged masses, and e = the error term.

I found the following results:

$$Y = 21.63 - 2.1 \ln X \qquad (2)$$

The correlation, r, the variance, r², and the probability, p, values are -0.97, 0.94, and 0.001, respectively. Thus, 94 percent of the variance or difference in planetary-rotation periods is explained by their masses. Since the value of p, 0.001, is much less than 0.05, the model is also statistically significant.

## Mercury and Venus: the Exceptions

The strong statistical relationship between the masses of the six planets and their rotation periods leads me to assume that such an outcome was not just a coincidence. Rather, the question that we need to pose and address should be why Mercury and Venus are outliers. Lack of moons is obviously one feature that distinguishes Mercury and Venus form their six other sister planets. When planets lack moons, one thing they might do is tidally couple their rotation periods to orbital periods of other objects like stars. For instance, Mercury is believed to be tidally coupled to the sun, since it rotates "not once per orbit but 1.5 times per orbit…" (Seeds, 2001: 472; see also Hughes, 2003). Thus, the slowness in rotational periods of Mercury and Venus may, in part, be explained by their lack of moons.

## The Kepler Problem

But why are planetary masses strongly related to rotation periods? There has to be a major reason. Without an explanation, the relationship between mass and rotation period would be just an empirical regularity. One of the most notable of such cases is Johannes Kepler's finding that orbital periods of planets and their distance from the sun are related. Kepler, the late 16th and early 17th century German astronomer, was unsuccessful in providing a scientific explanation (which Newton later found to be gravity) for the existence of such a relationship (Seeds, 2001). So, why are planetary masses and rotation periods related?

**Rotational Force or Torque**

The most plausible explanation for the existence of a relationship between planetary masses and rotation period is the presence of some kind of force. Indeed, it is a known fact in physics that torque is responsible for rotational motions of objects (Sears et al., 1982). Since planets rotate around their axes, they must be affected by torque. Given the importance of torque in rotational motion, I conducted further analyses on this subject. We know that the equation of torque is given by:

$$\Gamma = I \alpha \qquad (3)$$

where $\Gamma$ = torque, I = inertia, and $\alpha$ = angular acceleration.

We can see some interesting insights in Eq. (3). For instance, the acceleration of an object is proportional to the torque of the object. Thus, variation in acceleration of planets may explain the slowness of rotation periods of Mercury and Venus. In addition, the square of the radii of planets (since I = mr², where m = masses and r = radius) are proportional to their torques. These observations lead me to test the impact of acceleration and radius on planetary-rotation period. I calculated the acceleration data for the planets' rotation periods as follows. We know that angular acceleration is equal to translational or tangential acceleration divided by radius ($\alpha$ = a/R). We also know that translational or tangential acceleration, a, = V/T (velocity per time), and V = D/T, where D = distance; in this case, D is the circumference of the planets, $2\pi R$. Thus, the angular acceleration can by calculated by:

$$\alpha = 2\pi R/T^2 \qquad (4)$$

We also know that $2\pi R = 6.28$ radians. Using observational data for the rotation period, T, of the planets, I calculated the angular acceleration of the planets. As Table 2 shows, the angular acceleration data consistently vary (as is angular velocity) with the rotation periods of the eight planets. Thus, Venus and Mercury, the two tardy planets, have the slowest and second slowest angular accelerations, respectively.

I also ran a regression analysis to test the relationship between angular acceleration and rotation periods. As Eq. 5 shows, the relationship between angular acceleration and planetary rotation periods is linear, not logarithmic. The model shows that angular acceleration and rotation periods are inverse-linearly related, and the results are statistically significant.

**Table 2: Planetary Angular Acceleration, Angular Velocity, and Rotational Periods**

| Planets | Angular Acceleration (rad/hr$^2$) | Angular Velocity (rad/hr) | Rotational Periods (hours) |
|---|---|---|---|
| Venus | $1.84184441 \times 10^{-7}$ | $1.07548979 \times 10^{-3}$ | 5839.20 |
| Mercury | $3.17000807 \times 10^{-6}$ | $4.46179904 \times 10^{-3}$ | 1407.50 |
| Mars | $1.03605675 \times 10^{-2}$ | $2.55077173 \times 10^{-1}$ | 24.62 |
| Earth | $1.09639168 \times 10^{-2}$ | $2.62432094 \times 10^{-1}$ | 23.93 |
| Uranus | $2.11538338 \times 10^{-2}$ | $3.64480557 \times 10^{-1}$ | 17.23 |
| Neptune | $2.43785980 \times 10^{-2}$ | $3.91277259 \times 10^{-1}$ | 16.05 |
| Saturn | $6.00078927 \times 10^{-2}$ | $6.13880743 \times 10^{-1}$ | 10.23 |
| Jupiter | $6.38169875 \times 10^{-2}$ | $6.33064516 \times 10^{-1}$ | 9.92 |

$$Y = 24.85 - 870351\ X \qquad (5)$$

Moreover, the correlation, r, and the variance, r², are -0.94 and 0.89, respectively. Angular acceleration, however, has a bit lower r² than mass in explaining rotation periods.[3] Interestingly, the variance in planetary-rotation period that angular velocity explained is higher, 0.94, than that of acceleration, and the correlation between the former two variables is -0.97 (see Table 2 and Eq. 6).

$$Y = 32.58 - 27.1\ X \qquad (6)$$

In addition, I examined how planets' radii and their rotation periods are related. It is clear from Table 3 that planets' radii and their rotation periods are inversely related. However, despite its larger radius than Mercury and Mars, Venus has a longer rotation period. Moreover, despite Uranus' a bit larger radius than Neptune's, the former has a bit slower rate of rotation than the latter. I also ran regression to check the explanatory strength of radii on planetary-rotation period (see Eq. 7). The correlation, r, is 0.98 and the r² is 0.95. The model is also statistically

$$Y = -0.10 + 0.069\ X \qquad (7)$$

**Table 3: Planetary Radius and Rotation Period**

| Planets | Radius (km) | Rotation Period (hours) |
|---|---|---|
| Mercury | 2439 | 1407.07 |
| Mars | 3398 | 24.62 |
| Venus | 6052 | 5839.20 |
| Earth | 6378 | 23.93 |
| Neptune | 24750 | 16.05 |
| Uranus | 25559 | 17.23 |
| Saturn | 60330 | 10.23 |
| Jupiter | 71494 | 9.92 |

---

[3] This suggests that variance is not entirely explained by uniform movement of the independent and dependent variables but also by the rates of change between the two set of variables.

significant. The two variables are also positively and linearly related. Thus, each of the three components of torque - mass, acceleration, and radius - seem to have strong relationship with planetary-rotation period.

Finally, I obtained the torque values for each planet by using the equation ½m r²α for spherical bodies. I show this in Table 4. The data are consistent with previous analyses, that the higher the torque, the lower the planetary-rotation period. However, Venus, due to its larger mass than Mercury, has the second lowest torque.

I then ran a regression analysis to examine the effect of toque on planetary-rotation period (see Eq. 8). The torque data are logged. The correlation, r, is -0.98, r² is 0.97, and p is

$$Y = 18.66 - 0.85 \ln X \qquad (8)$$

0.0005. Torque explains 97 percent of the variance in planetary-rotation periods.

**Table 4: Planetary Torque and Rotational Period**

| Planets | Torque (m·N) | Rotational Periods (hours) |
|---|---|---|
| Mercury | $3.12091518 \times 10^{21}$ | 1407.50 |
| Venus | $1.64266752 \times 10^{22}$ | 5839.20 |
| Mars | $3.84242881 \times 10^{25}$ | 24.62 |
| Earth | $1.21022944 \times 10^{27}$ | 23.93 |
| Uranus | $6.00435808 \times 10^{29}$ | 17.23 |
| Neptune | $7.69070869 \times 10^{29}$ | 16.05 |
| Saturn | $6.21380037 \times 10^{31}$ | 10.23 |
| Jupiter | $3.09720862 \times 10^{32}$ | 9.92 |

The analyses so far suggest that planetary-rotation periods are strongly related to mass, radius, acceleration (or velocity), and torque. But, which of these variables are more important for planetary-rotation period? And which of these variables have a direct or indirect effect on planetary-rotation period? Before we answer the foregoing questions, let us identify equations that can help us calculate rotation periods. The makeup of these equations could shed some light on the variables that are more important in explaining planetary-rotation periods.

**Calculating Planetary-Rotation Periods**

Formulas for calculating planetary-rotation periods can be derived from torque, $\Gamma$, and angular momentum, L, equations. We know that torque and angular momentum are related by the equation, $\Gamma = dL/dt$. In other words, torque is the slope or the change in angular momentum with respect to the change in time. We also know that, $\Gamma = I\alpha$, and $L = I\omega$, where I = inertia (and I, in turn, $= mR^2$, where m = mass and R = radius). Assuming constant torque, $\Gamma = L/T$. By solving for T (the inertia, I, cancels out), we get:

$$T = -\omega / \alpha \qquad (9)$$

We can also derive T from the angular momentum equation, $L = I\omega$. Since $\omega = V/R$, and $V = 2\pi R/T$, we can solve for T. When the I (inertia) term cancels out, we are left with:

$$T = -2\pi / \omega \qquad (10)$$

The planetary-rotation period can even be calculated from the velocity-distance-time equation, $T = D/V$, where D = distance, in this case the circumference ($2\pi R$), and V = tangential velocity. Thus, T, assuming constant tangential velocity, is given by:

$$T = -2\pi R/V \qquad (11)$$

Thus, if we know a planet's angular velocity and angular acceleration (as in Eq.9) or its angular velocity (as in Eq.10) or its radius and tangential velocity (as in Eq.11), we can calculate its rotation period. Since rotation period and angular acceleration, angular velocity, and tangential velocity are inversely related, we will need to put negative signs on the right side of Eqs. (9, 10, & 11). The values of T are, however, always positive. In other words, we have to take the absolute value of the right side of Eqs. (9, 10, & 11). These equations also can be used to calculate the rotation periods of Mercury, Venus, and the Sun.

For instance, we can use Eq. (9) to calculate Mercury's rotation period. Since, from Table 2, $\omega = 4.46179904 \times 10^{-3}$ and $\alpha = 3.17000807 \times 10^{-6}$, T = 1407.504 hours.

We can also use Eq. (10) to calculate Earth's rotation period (see also Aoki et al., 1981). From Table 2, $\omega = 2.62432094 \times 10^{-1}$ rad/hour. T, is thus, = (2 x 3.14 rad) / 2.62432094 x $10^{-1}$ rad/hour = 23.93 hours.

Finally, we can calculate the Sun's rotation period by using any of the three equations. We know from observational data that the sun's radius and rotational velocity are $6.9599 \times 10^5$ km and $1.75 \times 10^5$ km per day, respectively. Its rotational period, using Eq. (11), is (2 x 3.14 x 6.9599 x $10^5$ km) / 1.75 x $10^5$ km/day = 25 days.[4]

## What to Make of All of These?

From what we have observed so far, velocity, acceleration, and radius seem to be the three most important proximate variables that can explain planetary-rotation periods. On the other hand, mass cancels out when we drive formulas of rotation periods from the equations of torque and angular momentum. This does not mean that mass and torque do not play any role in explaining rotation periods. Mass, for instance, influences the size of the radii of planets (r = 0.94); that is, the larger of the mass of a planet, the larger its radius. Mass also does influence the strength of the rotational force, torque. Indeed, the reason why radial distance, velocity, and acceleration are related to rotation period is because torque, heavily influenced by mass, may have played a major role in setting the initial constant speeds of planetary rotation. Interestingly, there seems to be a remarkable analogy between rotational and orbital periods of planets. First, radial distance, velocity, and acceleration seem to be the three most important and immediate factors in explaining both planetary revolution and rotation periods of planets. Second, torque and gravity, both heavily influenced by mass, seem to have played major roles in setting the initial constant rotational and orbital periods of planets, respectively. A major difference between the two types of planetary motions is that in revolution smaller radial distance between the sun and a planet means faster velocity and shorter orbital period. In rotation, a smaller-radial distance from the center of mass implies that a planet has a slower velocity and a longer spin period.

---

[4] Since this formula applies to the sun, it is also most likely applicable to the planets' moons that do spin.

## Conclusions

Using an inductive method of reasoning, this paper attempted to explain the factors responsible for variations in rotation periods of planets. The analyses in this paper suggest that planetary-rotation periods vary positively and linearly with their radii and are an inverse linear function of their velocities and accelerations. When rotation period is a function of the fraction of angular velocity and angular acceleration, however, it has a direct impact with the former variable. On the other hand, the effect of mass on planetary-rotation periods seems to be indirect: it influences the size of the radii of planets; that is, the larger the mass of a planet, the larger its radius. Mass also does influence the strength of the rotational force, torque, which may have played a major role in setting the initial constant speeds of planetary rotation...

There is one unanswered question, however: why are the rotation periods of Mercury and Venus longer than normal? I have indicated that it may have partly to do with their lack of moons. Although Park (2008) does not directly argue, as I do, that Mercury's and Venus' slowness has to do with their lack of moons, he seems to suggest that moons affect planetary spins. He states (2008:17), "Although the planets having satellites rotate much faster than …Mercury or …Venus, the fast rotations are attributed to the reaction torques by their satellites…" However, it should be noted that the presence of an orbiting moon around a planet is not likely the only reason for planetary rotation. Even moonless Mercury and Venus do rotate, albeit slowly. Rotational motions may have primarily to do with initial conditions of planetary formation including angular momentum. Thus, the effect of a moon on a planet is likely to be mediating the sustenance of planetary-rotation period. Conversely, the absence of moons may lead to several things including a synchronization of planetary rotation to other objects like stars or to a more than normal slowing process of planetary-rotation period.